\documentclass[preprint,prx,longbibliography]{revtex4-1}
\usepackage{graphicx}
\usepackage{comment}
\usepackage{amsmath,amssymb,amsthm} 
\usepackage{bm}
\usepackage{verbatim}
\usepackage{float}
\begin{document}

		\title{ The role of local bond-order  at crystallization  in a simple supercooled liquid}

\author{ S\o ren  Toxvaerd }
\affiliation{DNRF centre  ``Glass and Time,'' IMFUFA, Department
 of Sciences, Roskilde University, Postbox 260, DK-4000 Roskilde, Denmark}
\date{\today}

\vspace*{0.7cm}

\begin{abstract}
Large scale Molecular Dynamics simulations of sixty-five systems with $N$=80000 Lennard-Jones particles at
two  different supercooled liquid state points
reveal, that  the supercooled states contain  spatially heterogeneous distributed subdomains of particles with significant higher 
bond-order than the mean bond-order in the supercooled liquid.
The onset of the crystallization  starts in such an area with relative  high six-fold bond-order
for a supercooled state,
but  low bond-order for a fcc crystal state, and the crystallization is initiated by a nucleus where    all particles in the critical nucleus  have a significant
lower bond-order than in a fcc crystal. The critical nucleus of $N \approx 70 $ particles is surrounded by many hundred of particles  with relative high 
supercooled liquid bond-order and many of these particles are aligned with the crystal ordered particles in the critical nucleus.
The crystallizations are very fast and supported
by a fast growth  of the supercooled areas  with relative high liquid bond-order.
 The crystallizations  are to fcc crystals, but a significant
part of the crystallizations exhibit five fold
arrangements of polycrystalline subdomains mainly with fcc  crystal order and sign of hcp crystallites. 

\end{abstract}
\maketitle

\section{Introduction}

Computer  simulations have allowed us to determine the dynamics of crystallization.
Alder and Wainwright's \cite{Alder} pioneering simulations  of systems of hard spheres   revealed, that
a  system of hard spheres at high packing  fraction  crystallizes into an ordered state with fcc lattice structure.
Later,  computer simulations of a system of particles with the more realistic Lennard-Jones (LJ) potential \cite{Hansen}  not only confirmed the hard sphere result, but they were also in agreement
  with crystallizations of noble gases at low temperatures. Many later simulations of simple systems
verified the result, that the minimum free energy of a condensed system of simple spherical symmetrical particles is for the
fcc lattice arrangement \cite{Frenkel1,Pronk,frenkel2009,Errington}.

\begin{figure}
\begin{center}
\includegraphics[width=5cm,angle=-90]{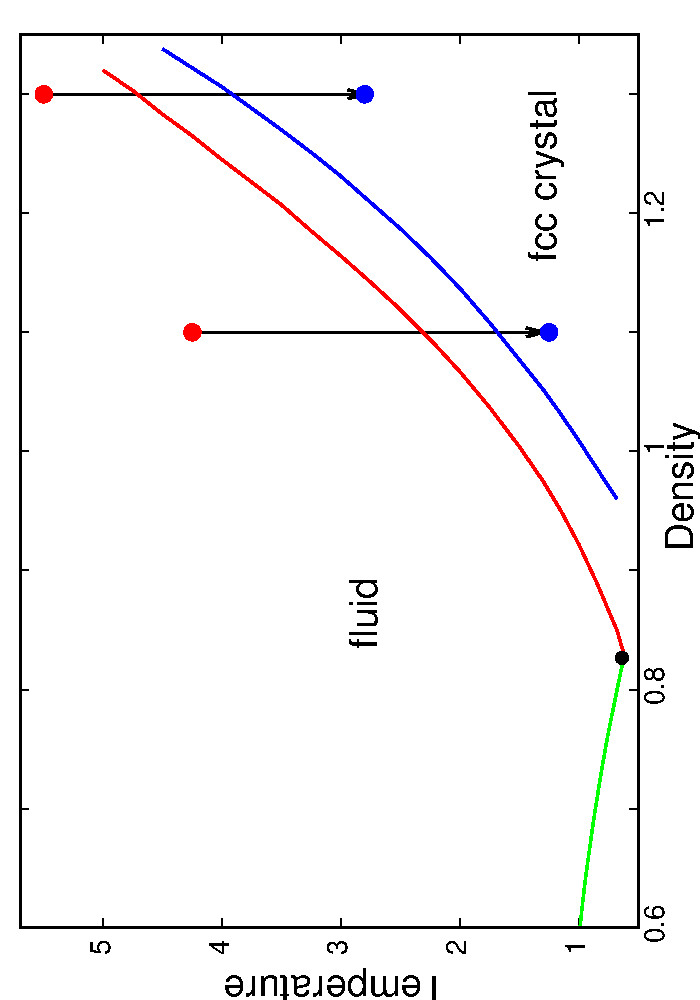} 
\end{center}
	\caption{The liquid-solid phase diagram for a LJ system. The
	liquid states,  which coexist with gas are with light green. The liquid states, which coexist with solid fcc are  red, and
	the coexisting fcc crystal states are blue. The MD systems are quenched from the liquids   (red points) down
	to the supercooled liquids (blue points). }
\end{figure}

 The theory of  crystallization is usually described by the classical nucleation theory (CNT)  and its extensions \cite{Volmer,Becker,Kalikmanov}.
 But an exact analyse \cite{Tanja1} and many computer simulations of nucleation of liquids as well as crystals in supercooled gases exhibit a much more complex
 dynamics with polymorphism \cite{Xu}, than given by CNT. The critical nucleus in a gas phase is not a compact object of the new phase, and
 it is not only initiated by density fluctuations, but also by temperature fluctuations  \cite{Toxa, Toxb}. A recent review of theory and simulations of crystallizations  is given in \cite{Jung}.

  The supercooled liquid exhibits bond orientational order \cite{Tanaka2012}, and
   here we show that  the bond-order, given by $Q_6$\cite{Nelson},  of the particles in a supercooled  LJ liquid is
    heterogeneous distributed. 
 The supercooled liquid  contains big  areas with significant higher bond-order than the mean bond-order in the liquid, and the crystallization
 is initiated from such an area. This result is consistent with the well known
 dynamic heterogeneities in supercooled liquids \cite{Ediger,Harrowell2004,Harrowell2006}.

 The crystallization is very fast  and accompanied by a growth of the supercooled areas with  relative high liquid bond-order.
 The  growing  crystal nuclei have  fcc bond-order, but in many cases the  nuclei also exhibit five fold symmetry
 with polycrystalline   domains. The  crystalline domains are mainly with fcc crystal arrangements, but some are also  with hcp structure. His behavior
 confirms previous results for homogeneous crystallizations in hard sphere systems
 \cite{Snook2003,Laso} and in a LJ system \cite{Daan,Medvedev,Delhommelle,Xu}.
 Many of the polycrystalline nucleations  ended in long time metastable polycrystalline states.

\begin{figure}
\begin{center}
\includegraphics[width=5cm,angle=-90]{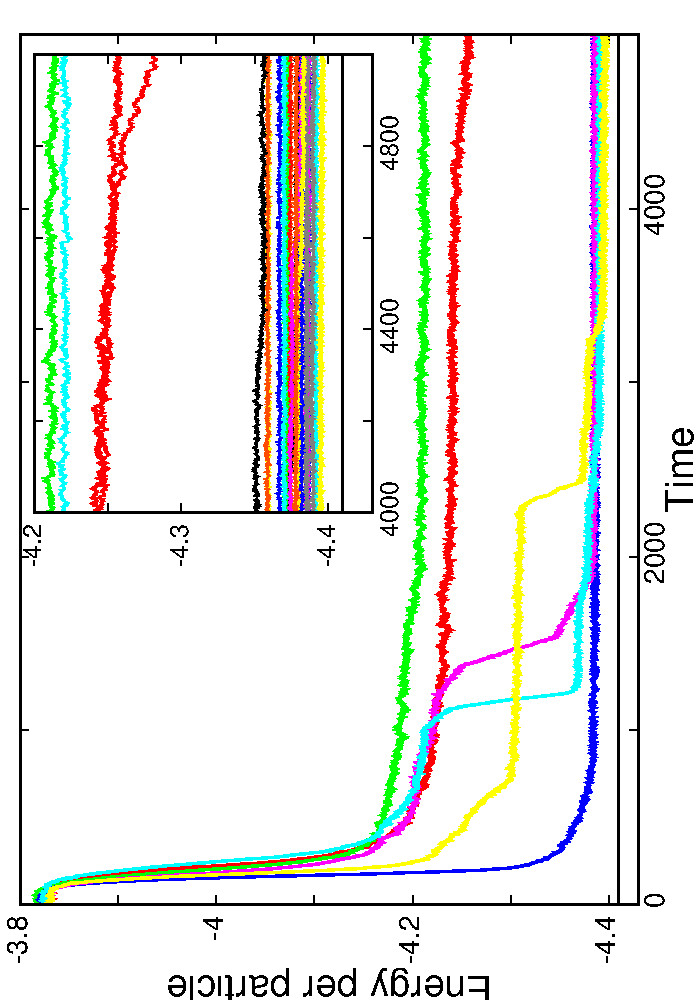} 
\end{center}
	\caption{Energy per particle, u(t), as a function
	of the time after  cooling to the supercooled
	state  $(T,\rho)$=(1.25,1.10) (left blue point in Figure 1). The figure shows
	six representative examples of $u(t)$.
	The inset are the energies for all twenty five experiments in the time interval $\Delta t \in [4000,5000]$
	(i.e for one million time s.jpg after four million time s.jpg). The black straight lines in the figure and
	the inset are the energy per particle for
	a perfect fcc crystal.}
\end{figure}

\section{ Homogeneous crystallization }

Molecular Dynamics (MD) systems of 80000 LJ particles (see Appendix) at different liquid state points  
  were cooled down to the  supercooled 
  states  (Figure 1).
  We have performed twenty five  independent $NVE$ simulations of supercooling and  crystallizations at the state point
   $(T,\rho)$=(1.25,1.10), and twenty $NVE$ and twenty $NVT$ simulations  at the state point  $(T,\rho)$=(2.80,1.30).
    The degree of supercooling given by the ratio between the supercooled temperature $T$ and
     the freezing point  temperature of the liquid $T_f$ is 0.54 and 0.60, respectively and
     the systems
     crystallized at the time  $\Delta t \approx 50- 150$  after the quenches. (Units are given in LJ units, see Appendix 1, and the time unit $\Delta t=1$ corresponds to
     1000 MD time s.jpg.)

The crystallizations result in a decrease in pressure and energy and
were essentially completed within a crystallization time of  $\Delta t \approx 50$.
But for some of the simulations
the crystallizations were, however, first completed after longer times. The energies per particle
for the  $NVE$ crystallizations at  $(T,\rho)$=(1.25,1.10) 
 are shown in Figure 2, and the energies for the twenty $NVT$ crystallizations are shown in Figure 9.

The description of the dynamics of the crystallization is divided into three subsections: The description of 
the supercooled  liquid, the onset of crystallizations, and finally the crystallization  and  the description of the crystal states after the crystallizations were completed.

\begin{figure}
\begin{center}
\includegraphics[width=5cm,angle=-90]{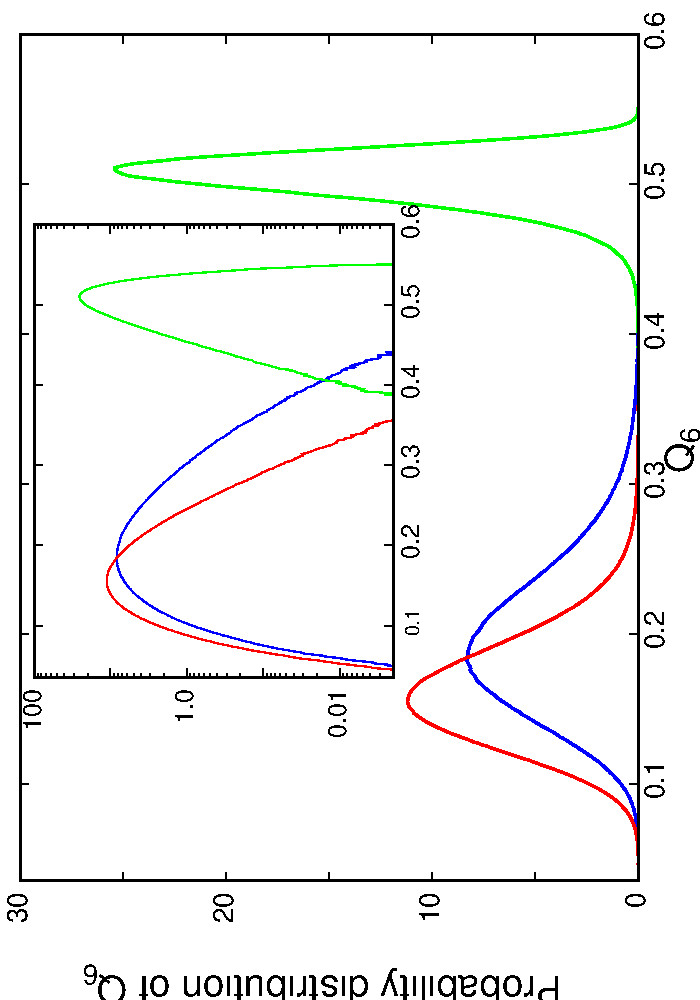} 
\end{center}
	\caption{Probability distribution of the bond-order parameter $Q_6$. The red curve is for the liquid before the 
	cooling (left red point in Figure 1) and the blue curve is for the supercooled liquid (left blue point in Figure 1).
	The green curve is for a perfect fcc crystal at the
	supercooled state point. The inset shows the $log$-distributions.}
\end{figure}

\subsection{ The supercooled liquid}
The bond-order function $Q_6$ (defined in Appendix 2) in the supercooled liquid is used to detect the dynamics of crystallization.
The liquid state is characterized by having a low value of bond-order compared with the bond-order in
the crystal state. In \cite{Russo} the authors used bond-order in the Gaussian core model, which is  a prototype for soft spheres,
to analyse the onset of crystallization. They 
found, that the crystallization occurs in precursor regions of high bond orientation order, and that the crystal which first nucleates is
the one which has the closest symmetry to the ordered regions in the supercooled state.
A later investigation of the bond-order in a compressed hard sphere fluid found, however, that the hexagonal ordering appeared simultaneously with the
density fluctuation at the onset of crystallization \cite{Tanja3}.

The present investigation shows, that the supercooled liquid contains large subregions of particles with relative high
$Q_6$ bond-order, and the nucleus which initiates the crystallization has only   a bond-order which is somewhat higher than  the order in the heterogeneous distributed
bond-order domains, but on the other hand have a lower bond-order than in the crystal.

\begin{figure}
\begin{center}
\includegraphics[width=7cm,angle=-90]{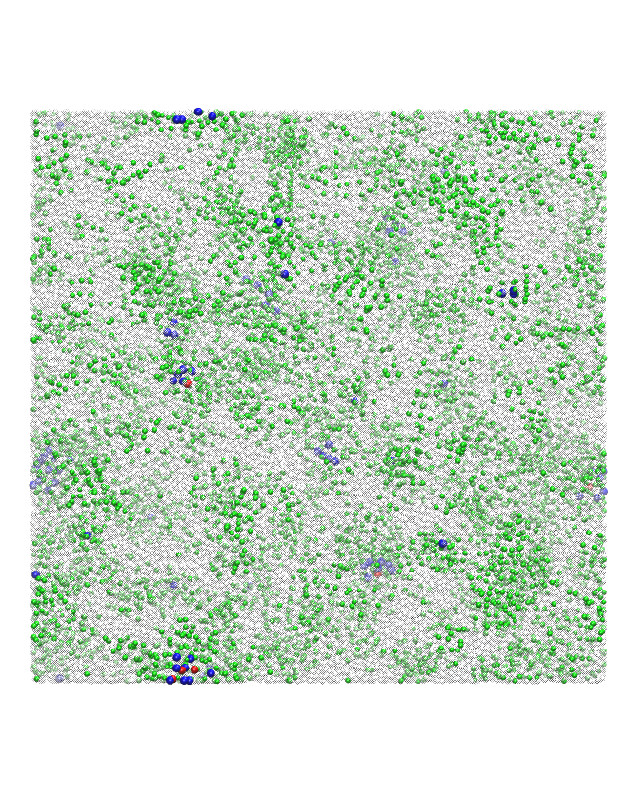}
\end{center}
	\caption{ Side view of the box with the $N$ particles.}
\end{figure}

\begin{figure}
\begin{center}
\includegraphics[width=5cm,angle=-90]{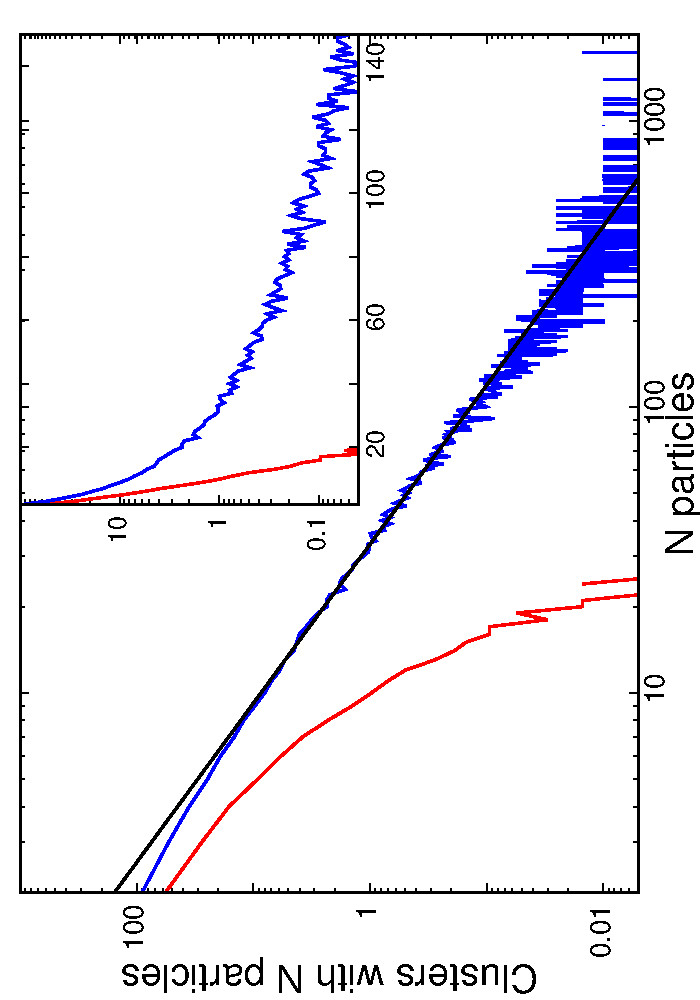}
\end{center}
	\caption{ Mean numbers  of clusters $n_{\textrm{N}}$ in the liquid states for clusters
	with N particles with bond-order $Q_6(i)>0.25$ for
	all particles $i$ in the clusters.
	The figure shows the $log(n_{\textrm{N}}(log(\textrm{N}))$ distributions and the inset is the $log(n_{\textrm{N}})(\textrm{N})$ distributions.
	Red curves are for the liquid (left red point in Figure 1) and blue curves are for the supercooled liquid.
	The straight black line in the figure is an algebraic fit, $a \times n_{\textrm{N}}^b$, to the distribution in the  supercooled
	liquid for clusters in the interval $n_{\textrm{N}} \in [20,200]$.}
\end{figure}

\begin{figure}
	\begin{center}
		\includegraphics[width=5cm,angle=-90]{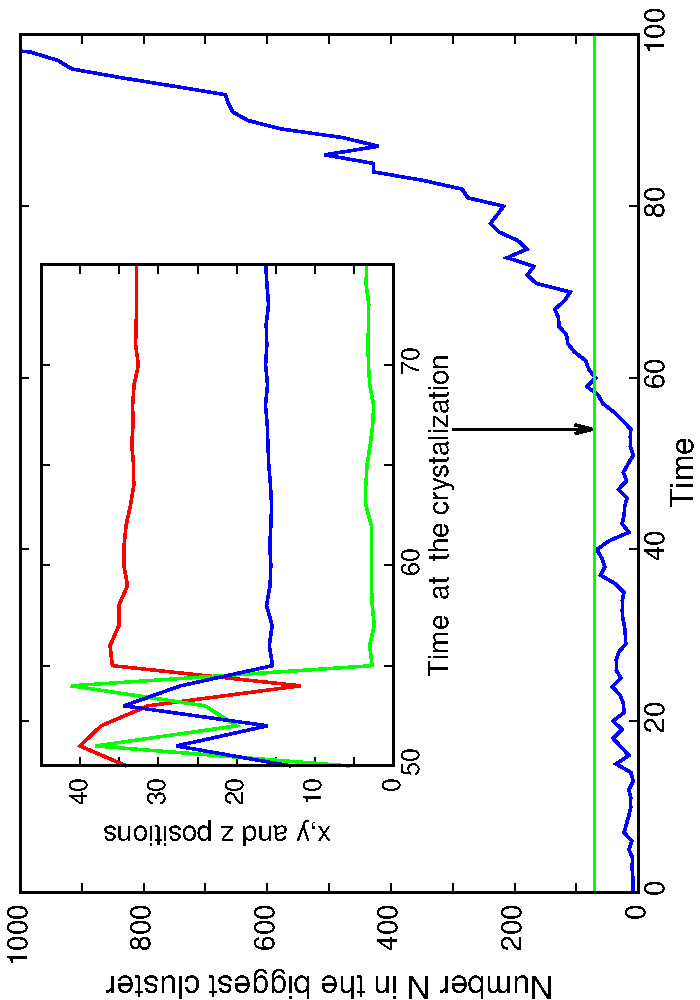}
	\end{center}
	\caption{ Number N$(t)$, in one of the twenty five simulations, of particles $i$ in the biggest  cluster 
	of crystal-ordered particles with 
	$0.35 < Q_6(i) $. The onset of crystallization at $t_{cr}$=56
	is marked by an arrow. From there the crystal nucleus  grows quite monotonically. The green line is an estimate of the critical crystal  cluster size of $N_{cr.} \approx 70$. The inset shows the x-, y- and z-positions of the biggest cluster.}

	 \end{figure}

\begin{figure}
\begin{center}
\includegraphics[width=8cm]{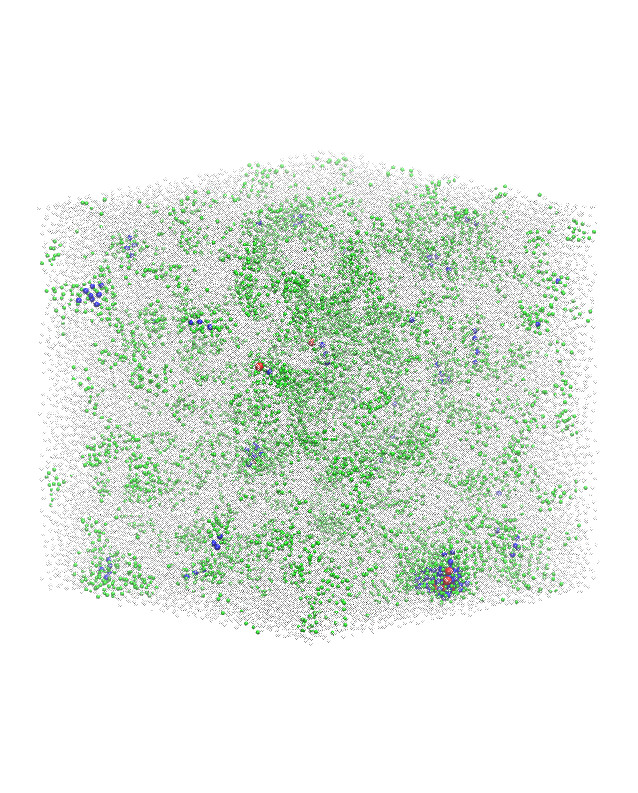}
\end{center}
\caption{ The particle positions at $t=0.58$  in the system with the $N(t)$ shown in the previous figure. The  local
environment with the critical nucleus is shown in the next figure. Color as  in Figure 4.}
\end{figure}

The distributions of the order parameter $Q_6(i)$ for particles $i$ in the liquid state   $(T,\rho)=(4.25, 1.10)$  (left red point in Figure 1) and
in the supercooled state   $(T,\rho)$=(1.25, 1.10)   (blue point) are shown in Figure 3 together with the distribution in 
a fcc crystal state at  $(T,\rho)$=(1.25, 1.10). The distributions for the supercooled
state (blue curve) and for the fcc crystal state (green curve) are separated, but
the $log$ distribution of  $Q_6(i)$ in the inset of the figure reveals, that there is an 
exponential decreasing probabilities for low order in the crystal state and a high order in the
supercooled state, and that there is an overlap of the two distributions in the region
$0.35< Q_6 <0.45$. Investigation of the crystal ordering at the creation of the critical nucleus (next subsection)
shows,  that the successful nucleus  have a mean order   $Q_6>0.35$ in accordance with the distributions in
Figure 3.
The  distributions of $Q_6$ for the liquid and the supercooled liquid, shown in Figure 3 are, however, different although
they have the same mean Gaussian-like shapes.  An analysis
of the distribution of particles with relative high bond-order reveal this fact.

From  the  locations of particles  with different values of $Q_6(i)$ for the particles $i$ in the supercooled liquid
one can see, that the distribution is non-uniform, and that there exists big subdomains with  relative high values
of $Q_6(i)$ for all the particles in the domain. The next figure shows this.  
 Figure 4 is a side view of the positions at a certain time of particles $i$  in the supercooled liquid.
 The white transparent spheres are particles with  $Q_6(i) < 0.25$, green spheres have   $0.25 < Q_6(i) < 0.35$, blue spheres
 have  $0.35 < Q_6(i) < 0.40$,
 and the red  spheres are particles in the supercooled liquid
 with a value of the bond-order $ 0.40<  Q_6(i) $. The blue and red particles are  particles with
 a lattice order, which is sufficient for crystal nucleation. (The values of $Q_6$ are obtained as time averages
 over thousand time s.jpg, but the heterogeneous distribution is also obtained from
 shorter and longer time intervals.)
 The positions of the particles is  not  uniformly distributed, but contains large
 areas with relative high bond-order.

The cluster distribution of
particles in the system with a certain quality, e.g. a high $Q_6$ value, can be obtained directly during the MD simulation and
without a significant increase in computer time for the big system by using the nearest neighbour list \cite{Toxa}.
The clusters of N particles  with $0.25 < Q_6(i)$ and the mean number $n_{\textrm{N}}$ of clusters with N particles was  determined
directly during a MD simulation. 
The number $n_{\textrm{N}}$(N), of a cluster  with N particles with a bond-order $0.25 <Q_6$ within a time interval $\delta t=0.1$ is shown in  Figure 5.
The  value $Q_6(\textbf{r}_i(t))$ of the bond-order for a particle $i$ at position $\textbf{r}_i(t)$ in the
supercooled liquid state fluctuates with time,  and the values of $n_{\textrm{N}}$ in the
figure are the mean for 200 independent distributions of clusters
with  $Q_6(i) > 0.25$, and where $Q_6(i)$ is obtained as the mean bond-order of
a particle $i$ within the time interval $\delta t=0.1$
($\approx$ one mean vibration within the  shell of nearest neighbours).
The figure shows the mean number $n_{\textrm{N}}$ of clusters of N particles with relative high bond-order
$Q_6(i)$ in the liquid state   $(T,\rho)$=(4.25,1.10) (red curve)
and the corresponding distribution of clusters in the supercooled state (blue curve). The two distributions are
 essentially different. The figure shows $log(n_\textrm{N})$ as a function of $log(\textrm{N})$ and the inset gives
 $log(n_{\textrm{N}})$  as a function of N. The distribution in the liquid state is exponentially declining
 (red  $\approx$ straight line in the inset), whereas the distribution for bigger clusters in the supercooled liquid
 is algebraic (blue $\approx$ linear function and black straight line in the figure). The black straight line
 is a line obtained from a fit of $a \times \textrm{N}^b$ to the $log(n_{\textrm{N}}(log \textrm{N}))$ distribution in the interval $n_{\textrm{N}} \in [20,200]$.
 The distribution of $n_{\textrm{N}}$ 
 shows also  that there   is regions of many  particles, as can be seen in Figure 4,
 with relative high liquid bond-order $Q_6 >0.25$. 

 The spatial  algebraic- or "heterogeneous" distribution of particles with significant higher bond-order than the mean order in the supercooled
 liquid is consistent with the well known dynamical heterogeneity in supercooled liquid \cite{Ediger,Harrowell2004,Harrowell2006}.
 In the next subsection it
 is stated, that the crystallization is initiated in such  a domain.

\subsection{The onset of crystallization}

The crystallization in a supercooled liquid appears when an ensemble of particles with lattice order gain   free energy
by increasing its size. In the CNT by, that the gain in free energy of the crystal phase exceeds the cost in surface free energy
by the increased surface of the crystal. In the MD ensemble simulation one primarily observes the crystallization, and the method gives not
a direct information about the free energy. For this reason it is not possible to  determine  the critical  nucleus precisely.
But one can locate the successful nucleus and its environment at the onset of nucleation.

The onset of crystallization is determined from the growth of the biggest cluster with bond-order $0.35 <Q_6$. In the supercooled
liquid the number N$_c$ of particles in
 the biggest cluster with  this crystal-like bond-order fluctuates with N$_c \le 100$ (Figure 6), but from the onset of crystallization 
the biggest cluster grows  very fast, as shown in the figure. The x-, y- and z- positions of the center of mass
are shown in the inset. The time record of these positions identify  the position of the successful nucleus even before it reaches the
critical nucleus size. The rather constant location of the center of mass of the biggest cluster at times $t \ge 56$, and the fluctuating
locations before $t=56$ show, that different crystal like nuclei appear in the system for $t < 56$, but
from then on the crystallization
is initiated by this nucleus. The size of the successful nucleus at $t=56$ is, however only $N_c=17$ and clearly much smaller than
the critical nucleus.  But after only 2000 time s.jpg the center consists of 68 particles with  $0.35 <Q_6$.
The green line $N=70$ is an estimate of the critical nucleus size, based on inspection of the twenty-five
$NVE$ simulation. As can be seen from Figure 6 some nuclei  at times $t< 56$ before the crystallization  sometime grow to this size.
This behaviour before the 
onset of nucleation is a typically
 for all  twenty five systems  in the supercooled states.

\begin{figure}
\begin{center}
\includegraphics[width=5cm,angle=-90]{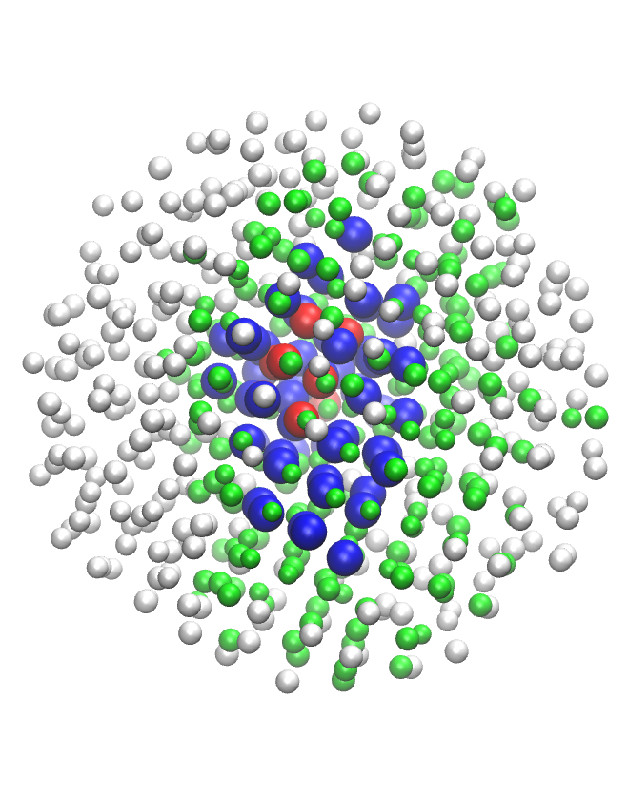}
\end{center}
	\caption{ The local environment of 574  particles within the sphere with radius 5 and with  the center 
	at the center of mass of the of the critical crystal nucleus  shown in Figure 7.
	 It consists  of N$_c$ = 68 particles
	with 59 (blue) particles with $0.35<Q_6(i)<0.40$ and  9 (red) particles with $0.40 < Q_6(i) $. The  225 small green particles
	have  $0.25<Q_6(i)<0.35$  and the small white transparent part particles have  $Q_6(i)<0.25$.}
\end{figure}

The positions of the particles at $t=58$ is shown in Figure 7 and with with the same color as in Figure 4. The critical nucleus is enlarged in the next figure.
Figure 8 shows the positions of the 574 particles within a sphere  with the center at the center of mass of the critical nucleus and with  the radius 5.
The critical nucleus consists of
59 (blue) particles with $0.35 < Q_6 < 0.40$ and 9 (red) particles with  $0.40< Q_6 < 0.45$ (no particle have a   $0.45< Q_6 $). There is 223 particles (green)
with  $0.25< Q_6 < 0.35$ and the particles with  $Q_6 < 0.25$ are  white transparent. The mean bond-order of the 68 particles in the crystal-like critical nucleus is 
$< Q_6 >$=0.38. The green particles are mainly located in a shell around the critical nucleus and with a tendency to fit into the lattice planes of the critical nucleus.
Even some of the  white particles with lower bond-order are located  in the critical crystals  planes.

The crystal bond-order $< Q_6 >$=0.38 in the critical nucleus is, however significant lower than the bond-order in a bulk fcc crystal,
 and the twenty $NVE$ and the twenty
$NVT$ simulations at the higher temperature and density $(T,\rho)=(2.40,1.30)$ show the same tendency.
All the critical crystal nuclei have a significant lower crystal  $Q_6$ value than a perfect fcc crystals.

The means for  the twenty-five crystal critical nucleus are\\
$ $\\
Mean   bond-order in the critical nucleus   $<Q_6(i)>=0.38\pm 0.02$\\
$ $,\\
at the number of particles in the  (estimated) critical crystal nuclei N$_{cr}=73 \pm 4$ \\

In conclusion: The  crystallization is initiated in a domain with excess bond-order for the supercooled liquid, but all the particles in
critical nucleus have a bond-order significantly less than the bean bond-order in the crystal (fcc) state. This result disagrees with the result in \cite{Tanja} for
homogeneous crystallizations in a system of hard ell.jpgoids.

$ $\\
\subsection{ The crystal states}

	\begin{figure}
\begin{center}
	\includegraphics[width=5cm,angle=-90]{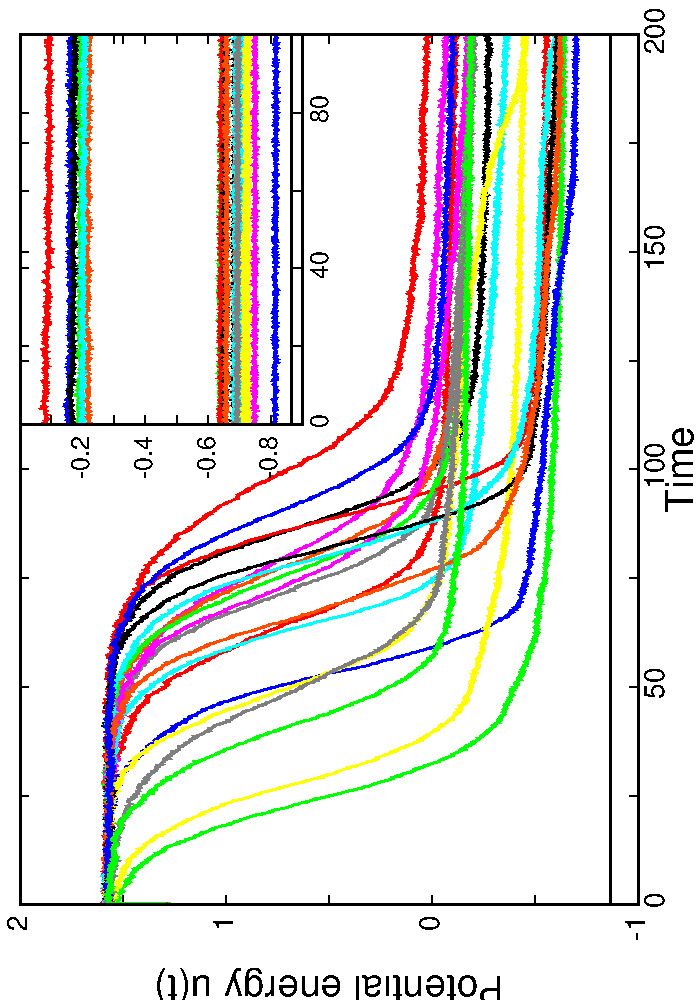}
\end{center}
		\caption{Potential energy per particle, u(t), as a function of time at
		the onset of crystallization for the ($T,\rho)=(2.80,1.13)$ (right blue point in Figure 1). The inset shows u(t) at the end
		of the simulations.}
	\end{figure}

Some crystallizations were completed within a  short time-interval of 50 to 100
time units, but  it took much longer times for many of the crystallizations as shown in
Figure 2.
Moreover, the systems did not end up in the same 
crystal state with energies close to the energy of a perfect fcc crystal. Some of the systems
ended up in states with  significant higher energies and pressures. For the twenty-five systems at the
relative low  density two to three of the simulations ended up in states with significant higher
energies than the other systems. This tendency is more pronounced for the twenty $NVE$ and the twenty $NVT$ simulations
at the higher temperature and density state $(T,\rho)=(2.80, 1.30)$.

Figure 9 shows the energy per particle u(t) for the twenty  $NVT$ simulations after the supercooling,
and with  the energies at the end of the simulations in the inset of the figure.
The energies  exhibit two energy bands, with fourteen of the energies  a little above the energy for a perfect fcc
crystal (black line), whereas there are six simulations that have  significant higher energies.  The same result was obtained for the twenty $NVE$
simulations at the same state point. From inspection of the particle positions and from the radial distribution functions it is is clear, that the fourteen
systems in the lower energy band are  fcc lattices with some defects,
but the crystal states in the systems with energies in  the upper energy band is more complex.

	\begin{figure}
\begin{center}
	\includegraphics[width=7cm,angle=0]{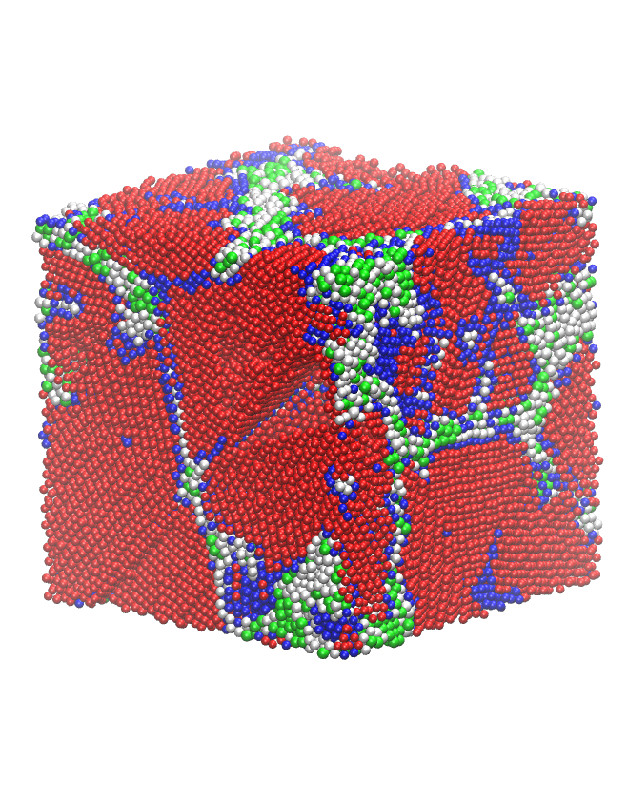}
\end{center}
		\caption{Side view of the particles in the system with the highest energy after the crystallization ( upper red curve   in Figure 9).
		The particles $i$ are colored as in Figure 4. With: low order
		with  $Q_6(i)<$ 0.20: green  $0.20< Q_6(i)< 0.30$; blue: $0.30< Q_6(i)< 0.45$;
		red:  $0.45 < Q_6(i)$.}
	\end{figure}

	\begin{figure}
\begin{center}
	\includegraphics[width=5cm,angle=-90]{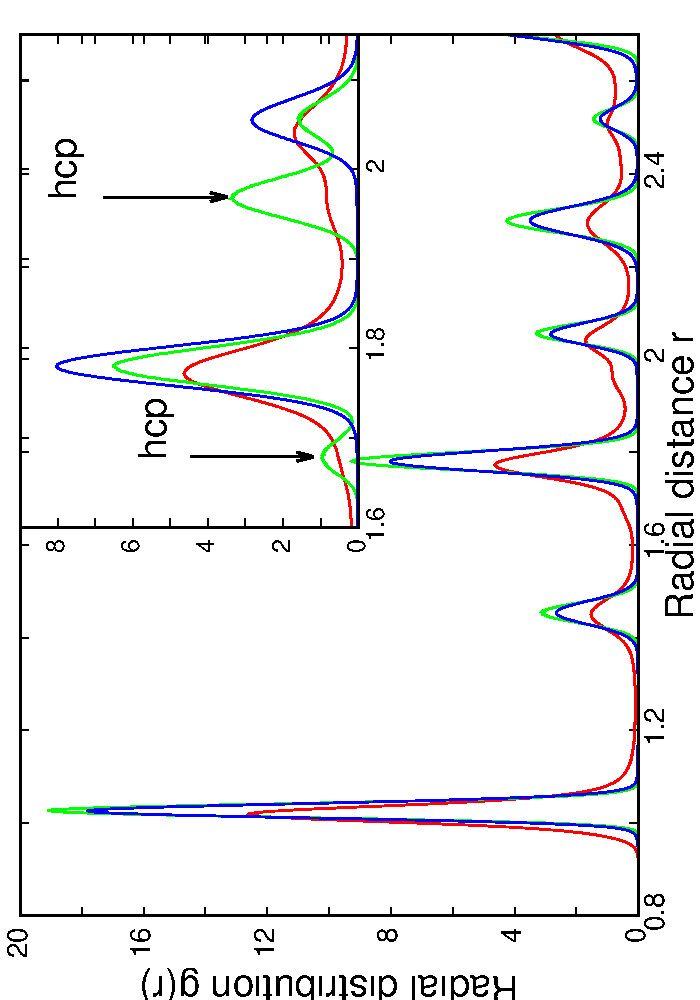}
\end{center}
		\caption{The radial distribution functions for the system with the lowest energy (blue)
		and with the highest energy (red) in the twenty $NVT$ simulations. After the crystallizations the systems were cooled down to $T=0.1$
		and compared with g(r) for a perfect fcc crystal (green). The two distributions
		are compared with a corresponding hcp crystal distribution (green) in the inset.}
	\end{figure}

The structures of the systems with energies in the high energy band are investigated in order to determine their lattice structure.
Particles interacting with  simple spherical symmetrical pair-potentials can exist in  many crystal arrangements \cite{frenkel2009}.
 But for a LJ system  at the two state point investigated here the fcc crystal structure has the lowest free energy \cite{Errington, frenkel2009}.
The positions
of particles for the system with the highest energy in Figure 9 (red upper energy function in the inset of Figure 9
is shown in Figure 10. The particles
are colored in accordance with their bond-order, and the colored positions show a complex crystalline structure with many areas with relative
low bond-order and even without
crystalline order ( $Q_6< 0.35$).

	\begin{figure}
\begin{center}
	\includegraphics[width=7cm,angle=0]{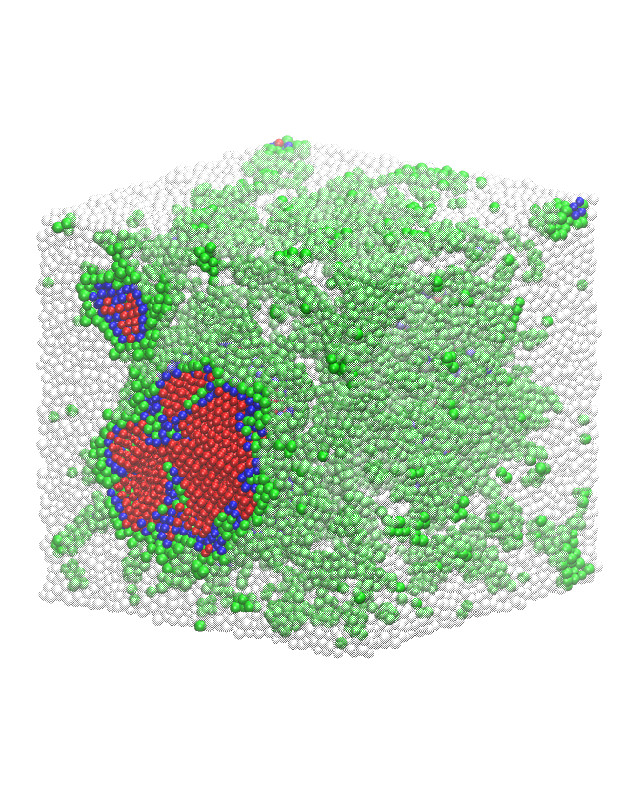}
\end{center}
		\caption{Side view of the particles  in the system (Figure 10), but at  a short time $\Delta t= 40$ after the onset of
		nucleation. The color of the spheres are the same as in Figure 10,
		but  particles with $0.20 < Q_6$ are   (white) transparent.
		The crystal cluster contains 5516 particles and the periodic boundary plane in front goes
		through the crystal nucleus.}
	\end{figure}

Crystalline order is traditionally determined from a systems structure factor or its
Fourier transform, the radial distribution function $g(r)$. The extremes in the  radial distribution functions at the high temperature $T=2.80$
are, however, not sharp due the particles vibrations  at their lattice positions. This thermal $``$noise" can be removed by cooling the system down to
a  low temperature. Figure 11 shows the radial distribution functions for the system with lowest energy (blue) and highest energy (red) in Figure 9,
and the distribution  $g_{\textrm{fcc}}(r)$  for a perfect fcc lattice (green), and after the systems were cooled down to $T$=0.1.
The figure confirms, that both systems mainly consist of particles in a fcc arrangement.
The disordered system with the highest energy deviates, however, from $g_{\textrm{fcc}}(r)$ especially at the radial distance $r \approx 1.9$.
The inset, where the two functions are compared with $g_{\textrm{hcp}}(r)$ (green) for a hcp crystal reveals, that the crystal (Figure 10) contains a small
number of particles with  hcp structure. The system with the highest energy was simulated over a very long time (47000 time units) in order to test whether
it is  stable, and it continued to be  in the polycrystalline state.

The polycrystalline state is ensured in the beginning of the crystallization, and it is not a final size
effect of the periodical boundaries. The next
figure shows the particle positions 40 time units after the onset of crystallization.
The crystal nucleus consists of 5516 particles and the periodic plane in front goes through the crystalline
nucleus and reveal, that the nucleus consists of
several small crystals with different orientations of the crystal planes. The green particles in the supercooled liquid with relative high bond-order percolate the 
system and a new crystal center  (upper left) is created with green particles included in the lattice arrangement. The figure shows that the spontaneous crystallization
is preformed also by growing  coherent order    in the supercooled liquid (green particles) , caused by the critical nucleus. This behaviour of the spontaneous
crystallization is found in all
the crystallizations.

\section{Conclusion}
 A supercooled liquid exhibits  spatially heterogeneous dynamics \cite{Ediger}, which influences the dynamic behaviour of the
 supercooled and   viscous liquid. Here it is determined, that the particles with relative high  bond-order in the supercooled Lennard-Jones system
 is heterogeneous distributed, and that the
 supercooled liquid contains  
 subdomains with  significant higher bond-order than the  mean bond-order in the supercooled liquid.
 Furthermore, the  crystal nucleation is initiated from such region with relative  high bond-order $Q_6$ for a supercooled liquid.
 The  domains with excess bond-order changes in extension and locations with  time,  and this is "consistent" with the dynamic heterogenities
 for supercooled liquid, but we have not directly established the connection between the domains of bond-order and the dynamics of the supercooled liquid, e.g. by determining the mobility of the particles in the subdomains.

  The  nucleus which initiates the crystallization is a relative  compact object with sixfold symmetry as assumed in the classical
  nucleation theory (CNT), but all the 
  $N\approx 70 $ particles in the nucleus have, however, 
   a significant smaller bond-order than  the bond-order in the fcc crystal. And in addition,  many of the surrounding particles  
   in the subdomain 
      are aligned with the particles in the initiating crystal nucleus (Figure 8).

   The growth of the critical nucleus is fast, and the crystal has percolate the
   big system of 80000 LJ particles  within a crystallization time of  $\approx 50$ to 100 time units.
   The systems did, however, not always end up in a homogeneous fcc state,
   but quite often they  ended up in a polycrystalline state. The polycrystalline state with 
   with traces of hcp crystallites
   is, however, established in the crystal
   nucleus shortly after the onset of crystallization (Figure 12). The polymorphism,
   where the particles crystallizes into different structures have already been observed  in LJ systems \cite{Delhommellejac,Delhommellejpc}.   
   Also here the areas with relative high bond-order in the supercooled part
   of the system plays a role for the fast spontaneous crystallization. The heterogeneous distributed areas grow fast and percolate
   the system  long times before  crystallization.

   The systems have been simulated by Molecular Dynamics $NVE$ and $NVT$ and with the same qualitative results. This is perh.jpg not
   surprising, because  there is only a marginal difference between the two MD methods for the big system. This fact is due to, that   the
   particles in the $NVT$ dynamics are constrained to a  thermostat temperature $T_{\textrm{th}}$ by the excess of kinetic energy of $\textit{the hole system}$.  The
   systems mean temperature $T$ in $NVT$  are smoothly constrained over longer times to the $NVT's$  $T_{\textrm{th}}$  value, which removes the latent heat during the
   spontaneous crystallization.
 .

				\section{Acknowledgment}
  Ulf R. Pedersen, Trond S. Ingebrigtsen  and Jeppe C. Dyre is gratefully acknowledged.
This work was supported by the VILLUM Foundation’s Matter project, grant No. 16515.

				\textbf{Author contribution statement}\\
				The articles idea, programs used in the simulations, the simulations and the writing are done by the author. 
	
\section{Appendix}
\textbf{1. The MD system and the computational details}

The system consists of $N=80000$ Lennard-Jones (LJ) particles in a cubic box with periodic boundaries, and the
crystallizations are obtained by Molecular Dynamics $NVE$ simulations with Newton's central difference algorithm \cite{Toxnew}
($``$Leap-frog").  The  time and length are given
 by the length unit $l^*=\sigma$ and energy unit $u^*=.jpgilon/k_{\textrm{B}}$
 in the Lennard-Jones potential for particles with the mass $m$. The unit of time
 is $t^*= \sigma \sqrt{m/.jpgilon}$.
 The LJ forces are
truncated and shifted at the interparticle distance $r_c=2.5$ \cite{tox1}, by which the system is energy stable
 \cite{tox2}. 
 The Molecular Dynamics is
 performed with a small time increment, $\delta t =0.0010$ due to the high densities in the supercooled liquids.

 The precise details of the phase diagram (Figure 1) for a LJ system depends on, from where the long range attractive forces
 are ignored, given by the value of $r_c$ \cite{tox4}. Most simulations of LJ systems including this one are 
 for  $r_c=2.5$,  for which the triple point densities are $(\rho_l,\rho_s,T)$=(0.8290, 0.9333, 0.618) \cite{tox5}. The present simulations
 of  crystallizations are for supercooled liquids at the state points
 $(T,\rho)$=(1.25, 1.10) and $(T,\rho)$=(2.80, 1.30) . The crystallizations 
   can be characterized as crystallizations in   supercooled  condensed liquid states.

 The  crystallizations are performed by cooling   from  the liquid states at  $(T,\rho)$=(4.25,1.10) and $(T,\rho)$=(5.25,1.30)
 down to the supercooled states  $(T,\rho)$=(1.25, 1.10) and $(T,\rho)$=(2.80, 1.30), respectively.
 The cooling  and $NVT$ simulations are  performed by a standard $NVT$ thermostat. The $NVE$ simulations are performed by
 cooling the high temperature system down  in teen thousand time s.jpg by the thermostat.  The $NVE$ supercooled state
 was accepted, if the  temperature in the succeeding ten thousand time s.jpg without 
 a thermostat  was within the temperature interval $T \in 1.25 \pm 0.005$.  
 The temperature in the systems increases at crystallization for  simulations without a thermostat. The $NVE$ systems were, however, so supercooled,
 that they ended up in crystal states (T,$\rho) \le $ (1.63,1.10) with total crystallization . A LJ fcc crystal at $\rho$=1.1  melts at
 T=1.68  \cite{Barroso}.

\textbf{2. Identifying crystal structure}

The crystal order is determined by a  modified bond-orientation order
 analysis \cite{Nelson,Dellago,Russo}.  A complex order parameter
\begin{equation}
	q_{lm}(i,t) \equiv 1/n_b \Sigma_{j=1}^{n_b}Y_{lm}(\textbf{r}_{\alpha \beta}(t))
\end{equation}
is calculated for each particle $i$ at time $t$. The sum over the $n_b$ particles $j$ in the local environment
runs over all neighbors $\alpha$ of particle $i$ plus the particle $i$ itself \cite{Dellago}. A potential neighbour $\beta$ to  $\alpha$ is
defined as a particle $j$ within the first coordination shell of  particle $\alpha$, given by the first minimum
in the radial distribution function. $\textbf{r}_{\alpha \beta}(t)$  is the vector between a particle $\alpha$ in $n_b$ and a nearest
neighbour $\beta$. The summation is further restricted to the sum over no more than
the twelve nearest neighbours \cite{Russo} (there are occasionally more than twelve nearest neighbours in the
first coordination shell in the supercooled liquid states).
The $Y_{lm}$ are the spherical harmonics, and the Steinhardt order parameter is defined as
\begin{equation}
	Q_l(i,t)=\sqrt{\frac{4\pi}{2l+1}\Sigma_{m=-l}^l \mid q_{lm}(i,t)\mid^2}
\end{equation}
In \cite{Dellago} the authors compared $Q_6$ and $Q_4$ for different crystal structures. The present LJ system
crystallizes into a fcc crystal, and  the best separation between  the bond-order in the
supercooled liquid states and in the crystal states is obtained for $Q_6$ (Figure 3). Finally,  a temporarily  
stable  crystal order at particle $i$ is determined by averaging over  a short time-interval of one time unit
\begin{equation}
	Q_6(i)=<Q_6(i,t)>.
\end{equation}

The clusters of particles with crystal order are obtained directly during the simulations \cite{Toxa}.  The threshold
value for crystal order is $Q_6(i)\approx 0.35$ accordingly to  Figure 3. A crystal nucleus is determined by, that 
all particles in the nucleus have an order  $Q_6(i)> 0.35$, and all particles in the nucleus have at least one 
nearest neighbour  particle $j$ with  $Q_6(j)> 0.35$. The cluster distribution is obtained directly during the simulations.
The biggest cluster is the successful nucleus, and  the center of mass of the biggest nucleus reveal whether there is
a competition between different growing crystals. It was never observed, once the successful  nucleus was established
(see inset in  Figure 6).

\end{document}